\documentclass[runningheads]{llncs}
\usepackage{graphicx}
\usepackage{enumitem}
\usepackage{xcolor}
\usepackage{pifont}
\usepackage{array}
\usepackage{multirow}
\usepackage{float}
\usepackage{makecell}

\usepackage{tcolorbox}
\usepackage{booktabs}   
\usepackage{ltablex}

\usepackage{url}

\usepackage{breakurl}
\usepackage[breaklinks]{hyperref}

\newcounter{researchquestion}
\newtcolorbox{researchquestionbox}[1][]{%
  colback=white, 
  colframe=black!75!black,
  fonttitle=\bfseries,
  title=Research Question \arabic{researchquestion}:,
  before upper={\stepcounter{researchquestion}},
  #1
}

\definecolor{mediumred}{RGB}{200, 0, 0}
\definecolor{mediumgreen}{RGB}{0,150, 0}
\definecolor{mediumyellow}{RGB}{220,180,0}

\newcommand{\BadEntry}[1]{\textcolor{black}{#1}}

\newcommand{\xmark}{\BadEntry{\ding{55}}}

\newcommand{\point}[1]{\par\smallskip\noindent\textbf{#1}. }

\begin{document}

\hyphenation{block-chain}

\title{SoK: Decentralized Sequencers for Rollups}

\author{Shashank Motepalli\inst{1,2} \and
        Luciano Freitas\inst{1,3} \and
        Benjamin Livshits\inst{1,4}
}

\institute{
    Matter Labs \and
    University of Toronto \and
    Télécom Paris, Institut Polytechnique de Paris \and
    Imperial College London
}
% % \authorrunning{F. Author et al.}
% % % First names are abbreviated in the running head.
% % % If there are more than two authors, 'et al.' is used.

\maketitle              % typeset the header of the contribution
\begin{abstract}
Rollups have emerged as a promising solution to enhance blockchain scalability, offering increased throughput, reduced latency, and lower transaction fees. However, they currently rely on a centralized sequencer to determine transaction ordering, compromising the decentralization principle of blockchain systems. Recognizing this, there is a clear need for decentralized sequencers in rollups. However, designing such a system is intricate. This paper presents a comprehensive exploration of decentralized sequencers in rollups, formulating their ideal properties, dissecting their core components, and synthesizing community insights. Our findings emphasize the imperative for an adept sequencer design, harmonizing with the overarching goals of the blockchain ecosystem, and setting a trajectory for subsequent research endeavors.

% Through our exploration, we ascertain that no existing solutions fully encapsulate all the desired properties. However, our delineated framework offers a foundation for the inception of such endeavors and establishes a benchmark for evaluating them as they evolve.

% Rollups have emerged as a leading solution to enhance blockchain scalability, offering increased throughput, reduced latency, and lower transaction fees. Yet, they are faced with a fundamental challenge: they currently rely on a single entity to determine transaction order, compromising the core decentralization principle of blockchain systems. Recognizing this, there is a clear need for decentralized sequencers in rollups. However, designing such a system is intricate. 

% In this paper, we (i) Identify the desirable properties of an decentralized sequencer; (ii) Explore the mechanisms and design choices essential for realizing these properties; (iii) Delve into unresolved questions and their implications for the system.
% In doing so, we show that there are currently no comprehensive solutions being implemented that achieve all the the desired properties, but the outline we define could pave the way to the stablishment of such projects and provide a way to compare them once they are more mature.

\keywords{Sequencers \and Rollups \and L2 \and Blockchains \and Decentralization}
\end{abstract}

\section{Introduction}

The rapid growth of blockchain applications has highlighted significant scalability issues. For instance, Ethereum processes approximately 12 transactions per second, leading to increased latencies and fees~\cite{mccorry2021sok}. To mitigate these challenges, off-chain scaling solutions are being developed to deliver high transaction throughput, reduced latency, and minimal fees. These ``off-chain systems", also called Layer-2 (L2) solutions, process transactions outside the base blockchain, referred to as Layer-1 (L1), while still leveraging the L1 to ensure security and integrity~\cite{kalodner2018arbitrum}.

Among off-chain systems, rollups stand out as a promising technology. They incorporate validating bridge smart contract on L1 that serves as an arbiter in the event of misbehavior or disputes~\cite{mccorry2021sok}. Rollups ensure high liveness with minimal fees and commit their state to the underlying L1. Based on their approach to ensuring integrity during state transitions, rollups can be categorized into two types. While optimistic rollups depend on fraud proofs for discrepancy detection, zk-rollups utilize validity proofs for discrepancy prevention~\cite{mccorry2021sok}. Other rollup variants, such as validium, rely on off-chain data availability~\cite{nazirkhanova2022information}. Despite their differences, a unifying component across all these rollup designs is the sequencer.

\point{Sequencers}
The sequencer is a pivotal component of rollups, responsible for ordering user transactions for execution. Currently, as illustrated in Table~\ref{tab:state-of-rollups}, sequencers are predominantly operated by centralized entities~\cite{l2beat}. This centralized design introduces significant liveness concerns, a challenge that is pervasive across all rollups. If the centralized operator becomes unavailable, the entire system's liveness is jeopardized and no transactions are committed. As of September~2023, Arbitrum One stands out as the only rollup with a deployed mechanism to address liveness issues, albeit one with its own set of flaws~\cite{l2beat}. Specifically, while it offers a process to replace a malfunctioning sequencer, this replacement is not only protracted, taking nearly a week, but is also restrictive, limiting the selection of a new sequencer to a pre-approved list~\cite{l2beat}. Other rollups currently lack any such mechanism, leaving them even more vulnerable to liveness disruptions.

\begin{table}[htp]
\caption{State of off-chain rollups as of August 2023. source: L2BEAT~\cite{l2beat}}
\centering
\resizebox{\textwidth}{!}{
\label{tab:state-of-rollups}
\renewcommand{\arraystretch}{2}
\setlength{\tabcolsep}{6pt}
\begin{tabular}{|l|c|c|c|c|}
\hline
\thead{Project} & \thead{Type of\\ Rollup} & \thead{Decentralized} & \thead{Sequencer \\ failure} & \thead{Force exit \\ (self-sequence)} \\ \hline \hline
Arbitrum One & Optimistic & \xmark & Live in a week & Via L1 smart contract \\ \hline
dYdX & ZK & \xmark & No liveness & \makecell{Via L1, requires\\ counter party to trade} \\ \hline
Optimism Mainnet & Optimistic & \xmark & No liveness & Via L1 smart contract \\ \hline
Polgygon zkEVM & ZK & \xmark & No liveness & No mechanism \\ \hline
Starknet & ZK & \xmark & No liveness & No mechanism \\ \hline
zkSync Era & ZK & \xmark & No liveness & \makecell{Via L1, but can\\  be refused} \\ \hline
\end{tabular}
}
\end{table}

Centralized sequencers also present the risk of monopolistic behaviors, such as transaction censorship or optimize their maximum extractable value (MEV). While these may not directly compromise safety by misappropriating user funds, still, funds could become inaccessible due to sequencer malfunctions or intentional censorship~\cite{mccorry2021sok}. Several rollups, including Arbitrum One, dYdX, Optimism mainnet, and zkSync Era, have introduced an ``escape hatch'' mechanism, allowing users to force exit the rollup via an L1 smart contract~\cite{gorzny2022ideal,l2beat}. However, current escape hatch implementations have limitations. For example, zkSync Era requires sequencer approval to include escape hatch transactions~\cite{l2beat}. Withdrawing funds from dYdX assumes the presence of a counterparty to trade, posing challenges during mass exits~\cite{l2beat}. Moreover, while most escape mechanisms address withdrawals, they do not cater to funds locked in a rollup's smart contract~\cite{gorzny2022ideal}. This underscores the pressing need for better sequencer designs.

\point{Contributions}
The problem this paper addresses is understanding the \emph{ideal properties} and responsibilities of decentralized sequencers. This concern is intrinsically tied to the scalability of L1 blockchains, thereby influencing the broader blockchain ecosystem. The problem's depth is underscored by a myriad untested community-proposed designs~\cite{b52,arbitrumdocs,starknetproposal}, indicative of a landscape rich in exploration. Exploring this domain is crucial, given the nuanced technicalities and hurdles associated with crafting robust decentralized sequencer specifications. 
We firmly believe that rollups should not merely emulate but indeed exceed the capabilities of their associated L1. 
In this paper, we offer a \emph{framework} for understanding decentralized sequencers in the context of rollups. 

The contributions of this work are three-fold: 
\begin{enumerate}
\item To the best of our knowledge, we are among the first to formulate the properties for sequencer in rollups, laying the groundwork for future research and development in this domain. 
\item We are also among the first to identify the key components of sequencers and elucidates their responsibilities.
\item Our work synthesizes community insights and best practices to present a holistic view for the rollup sequencer ecosystem.
\end{enumerate}

\point{Paper organization}
We begin with the background on the lifecycle of transactions within a rollup in Section~\ref{sec:workflow}.
In Section~\ref{sec:properties}, we formulate the ideal properties of decentralized sequencers along seven dimensions: functionality, decentralization, fairness, safety, performance, incentive design, and interoperability. We then delve into the components of a decentralized sequencer, in Section~\ref{sec:sequencerdesign}, namely: committee selection, sequencing policy, consensus mechanism, proposer selection, reward mechanism, data availability, interoperability, and governance. Drawing insights from community discussions, we identify a number of promising solutions and highlight potential shortcomings. As we conclude, we emphasize the challenges that researchers and developers must be aware of when devising decentralized sequencers in Section~\ref{sec:discussion} and Section~\ref{sec:conclusion}. We apply our framework on a decentralized sequencer proposal in Appendix~\ref{appendix:starknet}.

\section{Transaction Workflow in Rollups}
\label{sec:workflow}

Users interact with off-chain systems using client software, ranging from specialized wallets to command-line interfaces. Through these tools, users can initiate, sign, and send transactions, whether for asset transfers or to invoke functionalities within smart contracts.

\begin{figure*}[htp]
  \centering
  \includegraphics[page=1,width=\textwidth]{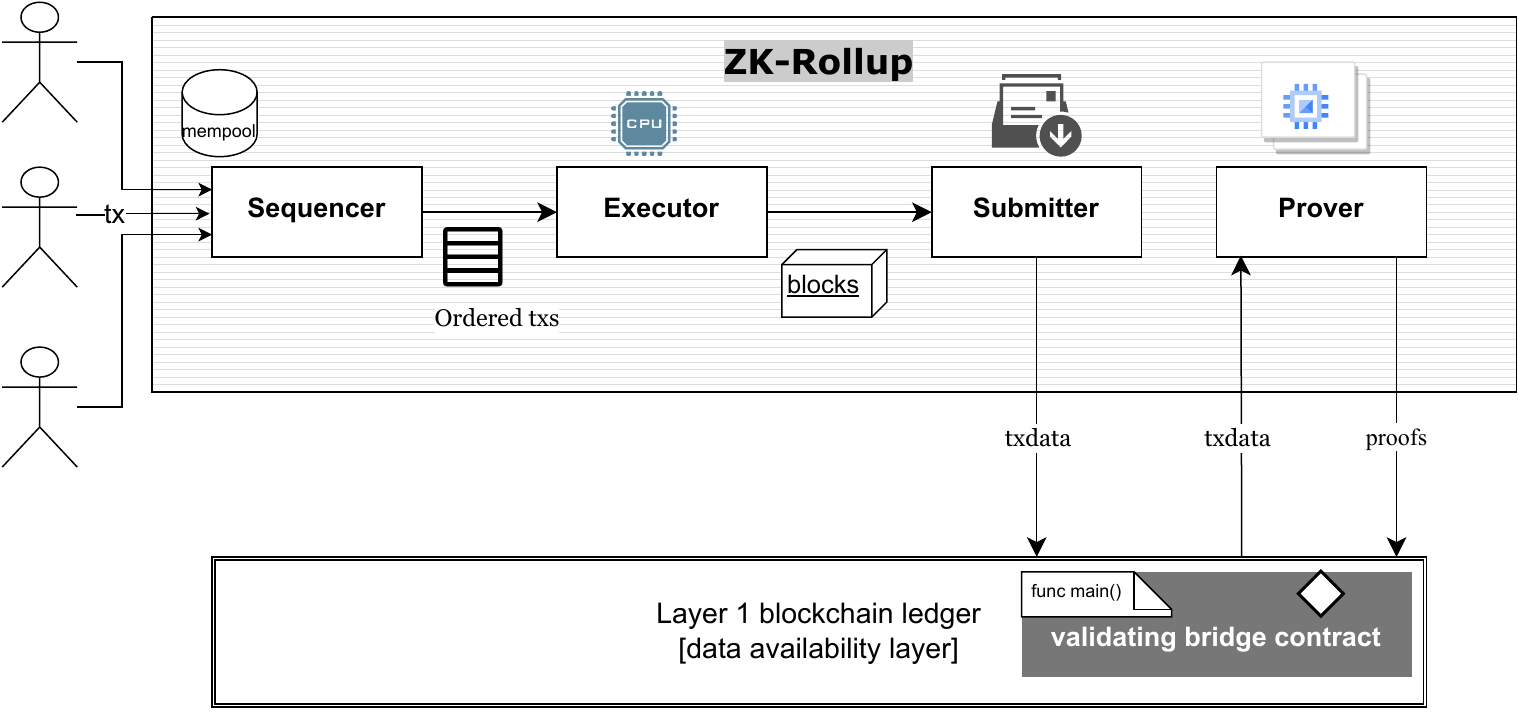}
  \caption{Transaction workflow in off-chain zk-rollup system that relies on L1 for data availability. txdata include merkle root and state differences.}
  \label{fig:zk-rollup-archiecture}
\end{figure*}

In this section, we focus on understanding the pivotal role of sequencers within rollup environments, using the transaction lifecycle in a zk-rollup as a reference~\cite{zksyncDocs}. A typical zk-rollup system includes four primary components, currently, all managed by centralized operator:
\begin{itemize}
    \item Sequencer: Collects, orders, and queues the pending user transactions.
    \item Executor: Processes transactions in the order provided by the sequencer.
    \item Submitter: Broadcasts the transactional data to L1 post-execution.
    \item Prover: Generates validity proofs for the finalization and validation of transactions.
\end{itemize}
For a detailed understanding, let us examine these components within the context of the transactional workflow, as depicted in Figure~\ref{fig:zk-rollup-archiecture}. 

Sequencers are tasked with collecting user transactions relayed from their clients. These sequencers maintain a memory pool of pending transactions, which is likely shared among a set of sequencers. Their main role is to order these transactions in preparation for execution, rather than directly creating blocks. Importantly, sequencers might remain unaware of the specific contents within a transaction, contributing to reduced latency of this step and, more importantly, potentially preserving the privacy of transactions.

Executors then take over, using the sequenced transactions to construct transaction blocks. It is important to note that block creation falls under the purview of executors, not sequencers. This design decision is influenced by potential L1 constraints that may become apparent only during execution. A salient example is the need to optimize for L1 gas fees during the block creation process. After formulating the blocks, executors process the contained transactions to revise the system state, typically represented in merkle trees. This operation culminates in a new merkle tree, the details of which, along with state differences, are handed over to the submitter. It's worth noting that merkle tree computations are notably CPU-intensive.

Submitters play their part by computing the hash of the merkle tree's root and forwarding the state differences to L1, leveraging the validating bridge smart contract. It is important to note that only the state differences, along with corresponding merkle tree root, are posted and not the transactions themselves~\cite{zksyncDocs}. This step is particularly resource-intensive due to the associated L1 gas fees. Submission can either be instantaneous post a block's creation or deferred until after a predefined number of blocks, with the timing often swayed by gas fee considerations. By performing these tasks, submitters effectively set checkpoints for the rollup blocks, offering robust \textit{data availability} guarantees.

In the final stretch, the provers actively monitor L1 checkpoint events, producing what are termed ``validity proofs"~\cite{goldreich1991proofs}. These proofs substantiate the computational integrity of state transitions to the new merkle root. This operation, though off-chain, is computationally demanding, warranting the use of GPUs. Once generated, these validity proofs undergo verification with minimal computational overhead on the validating bridge smart contract. The final step of committing these proofs to L1 signifies the attainment of complete transaction finality.

The distinction between rollup types becomes evident in this final phase. Unlike zk-rollups, optimistic rollups forgo validity proofs~\cite{mccorry2021sok}. Instead, they entrust the correctness of the state to external entities termed ``challengers." If these challengers discern inconsistencies in the proposed state by the submitter, they can flag them, providing fraud proofs that highlight the erroneous state transition~\cite{kalodner2018arbitrum}. Despite this fundamental difference, much of this paper, especially when focusing on sequencers, remains applicable to both zk-rollups and optimistic rollups.

% In practice, prominent zk-rollups such as zkSync Era, Starknet, and dYdX have predominantly centralized operators. This model promises a transition towards "progressive decentralization." In essence, roles such as the sequencer, executor, submitter, and prover, instead of being distributed, are typically centralized, sometimes even unified into a single operative entity. A similar centralized pattern is observable in prominent optimistic rollups like Optimistism and Arbitrium.

\section{Ideal Properties of Sequencers}
\label{sec:properties}

\begin{table}[!b]
\caption{Ideal properties of sequencers}
\label{tab:desirableproperties1}
\centering
\resizebox{\textwidth}{!}{
 \renewcommand{\arraystretch}{2}
 \setlength{\tabcolsep}{6pt}
 \begin{tabular}{|c|c|c|}
 \hline
 \multirow{4}{*}{\rotatebox[origin=c]{90}{\makecell{Basic \\ Functionality}}} 
 & Ordering transactions & Queues pending transactions, ensuring a coherent transaction flow\\ \cline{2-3}
 & Validity & Verifies transaction adherence to established rules of the underlying protocol \\ \cline{2-3}
 & Liveness & Timely processing of all valid transactions without indefinite delays \\ \cline{2-3}
 & Resilience & Ensures robust operation even amidst component failures \\ \hline
 \hline
 \multirow{5}{*}{\rotatebox[origin=c]{90}{Decentralization}}
 & No centralization & Does not rely on a singular centralized entity \\ \cline{2-3}
 & Permissionless & Allows honest parties to self-appoint without centralized authorization \\ \cline{2-3}
 & Collective decision & The decisions taken are influenced by most honest parties \\ \cline{2-3}
 & Sybil resistance & Thwarts attempts to generate multiple deceptive identities \\ \cline{2-3}
 & Geospatial distribution & Broad geographical spread of sequencers \\ \hline
 \end{tabular}
}
\end{table}

\begin{table}[!t]
\caption{Ideal properties of sequencers continued}
\label{tab:desirableproperties2}
\centering
\resizebox{\textwidth}{!}{
 \renewcommand{\arraystretch}{2}
 \setlength{\tabcolsep}{6pt}
 \begin{tabular}{|c|c|c|}
 \hline
 \multirow{4}{*}{\rotatebox[origin=c]{90}{Performance}} 
 & High throughput & Processes a vast number of transactions per second \\ \cline{2-3}
 & Low latency & Sequences with the least possible delay \\ \cline{2-3}
 & Fast finality & Swift, irreversible commitment for transactions \\ \cline{2-3}
 & Minimal resources & Minimal hardware requirements and reduced costs \\ \hline
 \hline
 \multirow{3}{*}{\rotatebox[origin=c]{90}{Safety}} 
 & Byzantine fault tolerance & Resilience against a predefined fraction of malicious sequencers within the set \\ \cline{2-3}
 & Data availability & Consistent access to transactional data for execution \\ \cline{2-3}
 & Rate limiting & Protects against targeted overload, including DDoS attacks \\ \hline
 \hline
 \multirow{4}{*}{\rotatebox[origin=c]{90}{Fairness}} 
 & Determinism & Predictable transaction order, free from arbitrary alterations \\ \cline{2-3}
 & Censorship resistance & All valid transactions are processed impartially, irrespective of origin/content \\ \cline{2-3}
 & Transparency & Mandates that all sequencing decisions are open, transparent, and verifiable \\ \cline{2-3}
 & Non-discriminatory fees & Fees are not manipulated to favor or penalize specific users/ transactions \\ \hline
 \hline
 \multirow{4}{*}{\rotatebox[origin=c]{90}{\makecell{Economic \\ Sustainability}}} 
 & Cost efficiency & Minimizes transaction costs for end users \\ \cline{2-3}
 & Incentive alignment & Aligns sequencers' rational behavior with system interests \\ \cline{2-3}
 & Reward distribution & Allocates rewards to sequencers fairly and transparently \\ \cline{2-3}
 & Slashing mechanisms & Penalizes malicious or non-compliant actions \\ \hline
 \hline
 \multirow{3}{*}{\rotatebox[origin=c]{90}{Interoperability}} 
 & Intra-ecosystem & Sequencing within rollups of a shared ecosystem \\ \cline{2-3}
 & Globally shared & Sequencing across various rollup ecosystems \\ \cline{2-3}
 & Cross-rollup atomicity & Atomic commitment across multiple rollups \\ \hline
\end{tabular}
}
\end{table}

Given that the idea of decentralized sequencers for rollups is early, we present a framework in Tables~\ref{tab:desirableproperties1} and ~\ref{tab:desirableproperties2} that outlines the ideal properties for sequencers. Each primary property is further dissected into specific sub-properties, providing a more granular perspective. While we suggest ideal properties, it is essential to understand that some properties might be mutually exclusive or vary based on the rollup's intended use. For instance, the requirements for an application-specific rollup could differ from those of a general-purpose rollup.

\section{Sequencer Design}
\label{sec:sequencerdesign}

In the previous section, we stated the ideal properties for an effective decentralized sequencer system. While these properties set the stage, a deeper dive into the system's core components is crucial for a comprehensive understanding. This section delves into each component. Our goal is to shed light on their primary objectives, inherent challenges, current best practices, and emerging designs from community discussions.

% \begin{figure*}[htp]
%   \centering
%   \includegraphics[page=1,width=0.6\textwidth]{figures/SequencerSystem.drawio}
%   % \includegraphics[width=0.6\textwidth]{figures/SequencerSystem.drawio.png}
%   \caption{Components of sequencer}
%   \label{fig:sequencer-components}
% \end{figure*}

It is important to emphasize that most of the designs we discuss are still in the developmental phase. The nuances of their implementation significantly influence their effectiveness and properties they satisfy. Our aim here is to give readers an overview of potential designs and their alignment with the properties we outlined. 

\subsection{Committee Selection}
\label{sec:committee}
\begin{center}
    \fbox{%
      \begin{minipage}{0.9\textwidth}
        \textbf{Motivation}: no centralization, permissionless \\
        \textbf{Consideration}: Sybil resistance, Byzantine fault tolerance \\
        \textbf{Impact}: resilience, transparency, slashing mechanism
      \end{minipage}
    }    
\end{center}
The committee selection component is pivotal in determining an optimal set of sequencers responsible for transaction collection and ordering within a rollup, directly influencing the system's integrity.

Currently, sequencer selection is predominantly centralized, with rollup operators maintaining exclusive control, as shown in Table~\ref{tab:state-of-rollups}. An exception is Arbitrum One, which employs a governance mechanism for sequencer transitions~\cite{arbitrumdocs}. However, its approach is constrained by a whitelist, with unclear inclusion criteria~\cite{l2beat}.

% The motivation behind a robust committee selection stems from the need for decentralization. A single sequencer system, while straightforward, is a central point of failure. A decentralized system, on the other hand, not only bolsters system resilience but also enables collective decision-making.

% Crafting a committee selection process that is transparent presents a myriad of challenges. At its core, the process should ensure that all stakeholders have a comprehensive understanding of the underlying sequencer selection criteria and mechanisms. An ideal system would be permissionless, opening the opportunities for any qualified entity to assume the role of a sequencer. However, this open-ended approach brings forth its own set of challenges. It necessitates the integration of robust mechanisms to thwart Sybil attacks and to ensure the system remains Byzantine fault-tolerant. Moreover, considering the geographical distribution of sequencers is essential to safeguard against potential localized system disturbances~\cite{motepalli2022decentralizing}.

\point{Sybil resistance}
To enhance the resilience of the system, it is imperative to select a set of sequencers rather than relying on a singular entity. While a permissionless selection process is the ideal, it inherently introduces vulnerabilities to Sybil attacks. Blockchains have proposed solutions such as Proof of Work (PoW), where participants use computational power to establish their identity~\cite{nakamoto2008bitcoin}. In contrast, Proof of Stake (PoS) involves participants staking tokens as a declaration of their identity~\cite{saleh2021blockchain}. Other notable mechanisms include Proof of Space~\cite{cohen2019chia} and Proof of Elapsed Time~\cite{motepalli2022decentralizing,corso2019performance}. Building upon Arbitrum One's approach, selecting a committee of sequencers through a transparent governance process emerges as a viable alternative~\cite{arbitrumdocs}. Given the evolution of DAO governance~\cite{sharma2023unpacking} and its potential for capital efficiency, this direction is worth further exploration~\cite{proofofgovernance}.

\point{Staking}
The prevailing trend in L2 sequencer committee selection is leaning towards staking~\cite{espresso2023,starknetproposal}. PoS offers several advantages. Firstly, by having a vested interest in the system, stakers are naturally driven towards non-malicious behavior. Secondly, PoS emerges as an eco-friendly alternative, requiring minimal computational resources. Its straightforwardness, coupled with a rich repository of existing knowledge, renders it a compelling choice. The potential for slashing in PoS addresses the ``nothing at stake" problem, ensuring active and correct participation~\cite{motepalli2021reward}. However, centralization remains a concern. As stakers accumulate more rewards, the risk of the system becoming increasingly centralized grows.

When adopting PoS for sequencer selection, the choice between using an L1 token or a native rollup token is crucial. An L1 token might be favored due to its association with gas costs and its widespread use in the community. However, a native token ensures rollup sovereignty and a more harmonized incentive structure.

\point{Restaking}
Rollups inherently derive their security from L1~\cite{mccorry2021sok}. Recent sequencer projects, including Astria~\cite{astriablog} and Espresso~\cite{espresso2023}, have proposed concept of ``restaking," where L1 validators are repurposed for sequencing tasks in the L2 domain. Utilizing the Eigen Layer~\cite{eigenlayer} allows these projects to tap into the security strengths of L1 validators. Yet, this approach isn't without its challenges~\cite{overloadEthereum}. The commitment of L1 validators to protect L2 environments also remains a point of contention.

A notable alternative to this approach is proposed by Starknet~\cite{starknetproposal} and Fernet~\cite{fernet}. Their key idea is to handle sequencer staking and committee selection through the corresponding L1 smart contract. This strategy preserves the sovereignty of sequencers in rollups while still benefiting from L1 security~\cite{starknetproposal}.

\begin{researchquestionbox}
What are the trade-offs and implications of different staking methods (e.g., native token, L1 tokens, restaking) in committee selection?
\end{researchquestionbox}

\point{Committee size}
Achieving the right balance in sequencer selection is crucial. While having too few sequencer nodes can lead to centralization, an excessive number can impede performance due to increased communication overheads required for consensus on transaction ordering~\cite{gavzi2023fait}. For instance, in the current Polygon network, the top~11 validators possess the authority to censor any transaction, and incapacitating the top 4 could bring the entire network to a standstill~\cite{polygondata}. Determining the ideal size for safety remains an open question. One way to manage an oversized sequencer pool is to select a smaller sequencer committee from the candidate pool, either through random methods~\cite{gilad2017algorand} or advanced deterministic functions~\cite{gavzi2023fait}. 

\point{Reconfiguration}
Regular reconfiguration of the sequencer committee is essential to replace faulty or malicious sequencers. Sequencers might either exhibit harmful behavior or experience crash failures due to routine maintenance~\cite{cohen2022aware}. Thus, it is crucial to schedule sequencer reconfigurations~\cite{blackshear2023sui}.

\begin{researchquestionbox}
How to determine the optimal sequencer committee size to prevent both centralization risks and performance impediments?
\end{researchquestionbox}

\subsection{Sequencing Policy}
\begin{center}
    \fbox{%
      \begin{minipage}{0.9\textwidth}
        \textbf{Motivation}: determinism, transparency \\
        \textbf{Consideration}: non-discriminatory fees, geospatial distribution  \\
        \textbf{Impact}: censorship resistance, Byzantine fault tolerance
      \end{minipage}
    }    
\end{center}
The sequencing policy delineates the rules governing transaction ordering. The sequencers must adhere to this policy. Notably, block construction in rollups diverges from traditional L1 mechanisms. Here, the primary objective is to optimize gas consumption on the underlying L1, necessitating a nuanced approach to sequencing policy design that accounts for these technical intricacies.

\point{FCFS}
Sequencers typically operate in a centralized manner, using the First Come First Serve (FCFS) strategy, as seen in rollups like Arbitrum One~\cite{mamageishvili2023buying}. In FCFS, transactions are timestamped upon reaching the sequencer, guiding their order of processing. While transaction arrival (inclusion) timestamp seems logical, it is vulnerable to ``latency racing". This is where users, aiming for faster transaction processing, position their systems close to sequencers to gain favorable timestamps~\cite{mamageishvili2023buying}.

\point{Maximal value}
Beyond FCFS, there are methods focused on maximizing block fees or transaction values. Given certain reward mechanisms, sequencers could be inclined towards this profit-centric sequencing. However, it's not without drawbacks. Chasing maximum profits, sequencers might deliberately postpone block creation, resulting in prolonged transaction wait times~\cite{mamageishvili2023buying}.

\point{Time boost}
This approach allows users to pay higher fees for a ``time boost", like a 5-second reduction in their transaction's timestamp~\cite{mamageishvili2023buying}. While beneficial for premium payers, its value diminishes with rising fees. This policy could inadvertently penalize users who choose not to or cannot afford to pay higher fees, creating a tiered transaction processing system.

\point{Fair BFT ordering}
Recent research underscores vulnerabilities in leader-based protocols, prompting the rise of ``order-fairness" in BFT systems when relying on timestamps~\cite{cachin2022quick,heimbach2022sok,kursawe2020wendy}. Protocols like Aequitas~\cite{kelkar2020order}, Pompē~\cite{zhang2020byzantine}, and Themis~\cite{kelkar2021themis} emphasize time-based order-fairness, enhancing transparency and robustness in decentralized consensus mechanisms against potential collusion and transaction manipulations. 

Furthermore, ensuring a geospatial distribution of sequencers can mitigate the advantages gained from latency racing, promoting a fair transaction processing environment~\cite{10237020}. 

\begin{researchquestionbox}
What are design considerations for sequencing policies to ensure fairness among sequencers, given latency racing and profit-driven delays?
\end{researchquestionbox}

\subsection{Consensus Mechanism}
\label{sec:consensus}
\begin{center}
    \fbox{%
      \begin{minipage}{0.9\textwidth}
        \textbf{Motivation}: collective decision, validity \\
        \textbf{Consideration}: Byzantine fault tolerance (BFT), high throughput \\
        \textbf{Impact}: resilence, fast finality, liveness
      \end{minipage}
    }    
\end{center}

The consensus mechanism ensures that all honest users agree on the same valid order of transactions~\cite{chaudhry2018consensus,gupta2019depth,zhang2022reaching}. Consensus mechanism offers two correctness guarantees:
\begin{itemize}
    \item {\bf Safety:} Given two transaction orderings output by two honest sequencers, one is prefix of the other.
    \item {\bf Liveness:} All valid transactions are eventually sequenced.
\end{itemize}

In the context of sequencers, the emphasis is on determining transaction ordering. Contrary to the state machine replication seen in L1s, sequencers eliminate execution requirements. This section delves into BFT mechanisms tailored for sequencers.

\point{Leveraging L1 for Consensus}
L1 can be harnessed for consensus in two distinct manners. The ``based rollups" approach~\cite{basedRollups} delegates the entire sequencing responsibility to L1. Here, L1's validators orchestrate the sequencing, interacting with the rollup smart contract to frame blocks with rollup transactions. Alternatively, sequencing can remain at the rollup level, but consensus among the parties processing the sequenced transactions is achieved via the L1 smart contract~\cite{aztecRFPResults}. 

Both strategies benefit from integrating L1's inherent safety and liveness attributes into the rollup. Employing L1 for consensus sidesteps the complexity of having separate consensus mechanisms for L2 sequencers and L1 validators. However, this method has its drawbacks. Depending on L1 for consensus might result in restricted block capacity and augmented latency, potentially offsetting the core benefits of rollups.
\point{Leader-based} 
HotStuff~\cite{yin2019hotstuff} represents a significant advancement in leader-based BFT consensus mechanisms for partially synchronous networks~\cite{castro1999practical}. It mandates a two-thirds node quorum for consensus and can endure up to a third of its nodes being faulty. Notably, HotStuff boasts linear message complexity, making the throughput depend only on the network latency, i.e., it is constrained on message delivery times rather than preset timeouts~\cite{malkhi2023lessons}. This was further optimized in HotStuff~2\cite{hotstuff2}, which reduced its latency from three to two round-trips. In the decentralized sequencer landscape, Espresso introduced HotShot, a modified HotStuff~2 for PoS sequencers~\cite{espresso2023}. 
% Weirdly enough I gotta put the footnote text here
% \footnotetext[1]{Ethereum blocks have per documentation target size of 15 million gas, but have a limit of twice this value. It is thus difficult to give a precise throughput value in terms of bytes because on top of variation of size in gas, there is also no direct relationship between bytes and gas. However, empirical data from~\cite{ethscan} indicate current average size $150~kB$, which we divide by block production time to obtain this number and contrast it with the Espresso results in the same metric.}

Tendermint~\cite{buchman2016tendermint,buchman2018latest}, on the other hand, has carved its niche in the consensus protocol landscape, powering a variety of L1s~\cite{tezos,cosmos,polygondata}. Tendermint shares many traits with HotStuff, including leader-based design, quorum size, and fault tolerance. The major difference is gossip protocol in Tendermint as compared to broadcast in HotStuff~\cite{buchman2022revisiting}. While the gossip protocol in Tendermint might not offer the same level of responsiveness as HotStuff, its design ensures a rapid and seamless leader transitions~\cite{buchman2022revisiting}. It also offers better workload distribution since all nodes need to send messages to only a small sample of nodes, irrespective of committee size $n$, ~\cite{amoussouguenou2019dissecting}, while a broadcast leader always sends $O(n)$ messages. Starknet~\cite{starknetTendermint} recently highlighted Tendermint as a potential candidate for their upcoming decentralized sequencer protocol, marking its relevance for sequencers.

\point{DAG-based} 
Keidar et al.~\cite{allYouNeedIsDAG} pioneered a consensus mechanism where every participant serves as a proposer. This mechanism necessitates the organization of proposals into a Directed Acyclic Graph (DAG), by requiring a proposer to refer to all previous proposals it has come across in the system, forming a data structure that represents the causal history of the proposals. This innovative approach fostered a fresh avenue of research, a significant milestone of which is the inception of the Narwhal mempool~\cite{narwhal}. This development delineated the roles of data dissemination and sequencing, a modification that not only amplified the protocol's throughput but also enhanced the efficiency of DAG construction. Protocols such as BullShark~\cite{bullshark}, BBCA~\cite{bbca}, and Shoal~~\cite{shoal} further advanced this space conferring this family of protocols high throughput and low latency. 

DAGs are worth exploring for L2 sequencers~\cite{starknetproposal} and are used in L1s such as Sui~\cite{sui}. Specifically, their Lutris~\cite{lutris} protocol harnesses \emph{commutativity} of some transactions in order to minimize what needs to be subjected to consensus. 

\begin{researchquestionbox}
    How do leader-based consensus mechanisms compare with DAG-based approaches in the context of decentralized rollup sequencers? \\
    Additionally, can L2 consensus leverage commutativity?
\end{researchquestionbox}

% It currently deployed in Layer-1 in project such as Sui~\cite{sui} use consensus protocol called Bullshark~\cite{bullshark} that runs on top of the Narwhal Mempool indicating how blocks should be ordered given the state of the DAG.

% One of the possible ways we could see DAG-based consensus mechanisms being introduced into L2 projects in the future is by Chainlink's Fair Sequencing Service (FSS)~\cite{chainlinkFSS} which counts Arbitrum as one of ecosystem members. Their research team has been making significant strides with DAG-based consensus aiming to reduce its latency~\cite{bbca} by changing the underlying broadcast primitive used to construct the DAG and incorporation of other properties demanded by blockchain applications such as MEV reduction~\cite{malkhi2022maximal}, as we shall discuss later. 
\subsection{Proposer Selection}
\begin{center}
    \fbox{%
      \begin{minipage}{0.9\textwidth}
        \textbf{Motivation}: ordering transactions, determinism, censorship resistance \\
        \textbf{Consideration}: no-centralization, collective decision\\
        \textbf{Impact}: rate limiting, rewards mechanism
      \end{minipage}
    }    
\end{center}
In consensus mechanisms, proposers—often termed leaders (Section~\ref{sec:consensus})—coordinate the process, such as signature aggregation on broadcast proposal in HotStuff~\cite{yin2019hotstuff}. Proposer selection (or leader election) mechanism in L2 protocols can underpin the native L2 consensus or be utilized by the L1 smart-contract to select from L2 sequencer committee.

Traditionally, consensus mechanisms had fixed proposers, initiating a view-change only upon proposer faults~\cite{castro1999practical}. Modern systems, however, emphasize frequent view-changes to satisfy chain quality~\cite{garay2015bitcoin}, promoting better load-balancing and censorship-resistance~\cite{lessonsHotstuff}. We now see mechanisms for proposer selection.

\point{Round-robin}
Round-robin is a straightforward proposer selection method, cycling through sequencers in a predetermined order~\cite{chan2020streamlet,yin2019hotstuff}. Its strengths lie in its simplicity and inherent fairness, ensuring each participant an opportunity to propose. It ensures deterministic termination, as there is a defined upper limit on the time until a correct leader emerges during network synchrony. Additionally, it guarantees uniqueness, there is a single proposer at any given time.

\point{Random Sampling}
Another prominent mechanism for selecting proposers with uniqueness guarantees. We map set of numbers to identity of sequencers and randomly generate a number from the set to choose proposer~\cite{bunz2017proofs}. This mechanism is the currently adopted in Ethereum~\cite{ethereumLeader} and is highlighted of L2 proposals such as Starknet's~\cite{starknetproposal} and Espresso's~\cite{espresso2023}. Among its advantages, it protects the system from bribery attacks, where a proposer can accept offers based on the position it occupies in the protocol; it also guarantees some protection against DDoS attacks if the time between the leader's identity reveal and the transmission of its proposal is short; finally, it incentives participants to remain available as they could be elected to the proposer role at any point.

\point{Cryptographic Sortition}
Algorand~\cite{gilad2017algorand} employs a randomness mechanism for proposer selection through cryptographic sortition. Each party utilizes a Verifiable Random Function (VRF)~\cite{micali1999verifiable} to locally ascertain their proposer status. Upon readiness, they broadcast their proposal alongside their VRF's verifiable result, enabling peers to validate their selection. This enhances DDoS resilience, though it may occasionally result in multiple or no proposers. Fernet~\cite{fernet}, a decentralized sequencer selection proposal for Aztec, adopts this approach.

\point{Single Secret Leader Election (SSLE)} 
In an SSLE scheme~\cite{boneh2020single}, a group of sequencers register to participate in a series of elections. Each election chooses exactly one proposer. The proposer knows they were selected, but others only learn their identity once the proposer reveals themselves, along with proof of their selection. SSLE scheme is gaining interest from both Ethereum~\cite{ethereumSSLE} and rollups~\cite{whisky}. 

\point{Reputation Mechanism}
All the discussed proposer selection mechanisms can falter if a faulty party is repeatedly chosen, causing system delays. To mitigate this, historical system observations can inform a reputation-based selection, bypassing non-performing parties. While various proposals exist~\cite{cohen2022aware,shoal,bbca}, the challenge lies in avoiding faulty parties without inadvertently censoring others.

\point{Proposal scoring} 
While previous approaches pre-select a proposer, minimizing the cost of receiving redundant proposals that will be discarded, an alternative is to allow all sequencers to propose. This method selects the proposals based on a predefined scoring criterion, motivating block production aligned with specific goals such as maximum extractable value. For instance, Aztec's B52 sequencer selection proposal~\cite{b52} permits provers to vote on preferred blocks, optimizing system profitability. We note that cryptographic sortition can be seen as a particular case of proposal scoring where the score is given by a VRF.

\begin{researchquestionbox}
    Could we modify protocols originally intended for a single-proposer model to incorporate a hierarchy of proposers?
\end{researchquestionbox}

\subsection{Reward Mechanism}
\begin{center}
    \fbox{%
      \begin{minipage}{0.9\textwidth}
        \textbf{Motivation}: incentive alignment, reward mechanism\\
        \textbf{Consideration}: determinism, cost efficiency\\
        \textbf{Impact}: transparency, slashing mechanism
      \end{minipage}
    }    
\end{center}
Sequencers are responsible for the collection and sequencing of transactions. Ensuring their appropriate compensation is crucial for the economic sustainability in decentralized systems~\cite{catalini2020some}. Such rewards should not only incentivize honesty but also be distributed in a manner that is transparent, deterministic, and fair.

\point{Costs} 
Let us begin by understanding the assiociated costs linked to rollups. Currently, centralized operators bear these costs~\cite{l2beat}. The main expenditure is L1 gas fees for transaction submissions~\cite{rollupeconomics}. With volatile L1 token prices, especially if L2 fees are collected in native token, submitters might delay posting L2 rollup transaction state data batches on L1 to benefit from the price fluctuations~\cite{mccorry2021sok}. This economic design challenge is  modeled using a Markov decision process~\cite{mamageishvili2022efficient}.

Other potential costs vary by design. For example, zk-rollups require significant computational power for proof generation. Depending on the design, there might be costs for data availability layers, especially if rollups opt for third-party data availability layers. There are also capital costs in PoS systems and bonding costs for challengers in optimistic rollups. To ensure system viability, rewards should cover these costs. With multiple L2 operators, reward distribution becomes a collective endeavor~\cite{boldarbitrum}.

\point{Fees and staking} 
Sequencers can primarily earn through transaction fees, which users pay to have their transactions included. While transaction fees are a prevalent form of reward, it is essential to consider subsidies for other system components~\cite{fanti2019economics}. Staking rewards are a popular design choice, but they come with challenges. Those who opt not to stake might face penalties, potentially leading to an imbalanced system~\cite{proofoffee}. Moreover, this leads to overcompensation. It is crucial to recognize the distinct roles of sequencers in comparison to L1. Unlike L1, sequencers are tasked solely with ordering transactions. They don't handle execution, proof generation, or L1 updates, all of which are more resource-intensive and costly~\cite{rollupeconomics}.

\point{Contract-based compensation} 
Another intriguing model views sequencers as contractors, receiving fixed payments for their services. In this setup, the system compensates elected committee members equally for their transaction ordering duties without capital locking. An extension to this idea is an auction mechanism where sequencers bid by specifying their required incentives. Upon exhibiting honest behavior, their rewards would be the total reward minus their bid~\cite{proofoffee}. Such models seem to align more closely with the unique requirements of L2 sequencers and warrant further exploration.
\begin{researchquestionbox}
    How effective are contract-based compensation (or auction) models in comparison to traditional reward structures for sequencers? 
\end{researchquestionbox}
\point{Exploring rollup specific designs} 
One unique approach tailored for zk-rollups involves provers compensating sequencers. This could lead to delays based on when provers feel it is viable to generate proofs~\cite{tsabary2018gap}. However, this design might foster collusion between sequencers and provers. There is a risk that they could act in self-interest, prioritizing certain provers over others.

\point{Reward distribution considerations} 
Having explored the sources of rewards, the focus now shifts to their distribution. The central question is whether rewards should be distributed equally or be proportional to stake~\cite{motepalli2021reward}. Furthermore, when considering the distribution between the sequencer and the rest of the participants in rollups such as executors, submitters, and challengers or provers~\cite{rollupeconomics,mccorry2021sok}. Their collective efforts should be acknowledged uniformly, one could create contracts for the same. 
\begin{researchquestionbox}
    How can we design comprehensive economic models that capture the various costs, rewards, and dynamics of L2 systems?
\end{researchquestionbox}
\subsection{Data Availability}
\begin{center}
    \fbox{%
      \begin{minipage}{0.9\textwidth}
        \textbf{Motivation}: data availability, Byaznatine fault tolerance\\
        \textbf{Consideration}: high throughput, low latency, cost efficiency\\
        \textbf{Impact}: no centralization
      \end{minipage}
    }    
\end{center}
The Data Availability (DA) layer is pivotal in decentralized systems, ensuring consistent access to transactions data~\cite{al2019lazyledger}. In L2 contexts, the DA's role is accentuated due to the trust-minimized environment as compared to L1s. This necessities the need for \emph{full nodes} with synchronized system views~\cite{sui}. Integrating DA layers in decentralized sequencers can enhance throughput and lower communication costs by ordering short digests instead of directly referring to individual transactions~\cite{narwhal}. 
One of the most important considerations about DA is where to store data.

\point{On-chain}  
Some rollups opt to store the data in the L1 blocks published as call-data. The main advantage  is that the rollup inherits full security and censorship resistance of the base blockchain. On the other hand, it has the downside of incurring huge costs as storing data on L1 is expensive. Danksharding~\cite{dankshardingBlog,danksharding} offers a solution by attaching temporary data blobs to blocks, which are deleted post-verification, ensuring efficient processing without permanent storage overheads.

\point{Off-chain without replication} 
The cheapest and least secure form of DA is generally called \emph{validium} where a single off-chain node stores all the data necessary for treating transactions is held by a single party~\cite{nazirkhanova2022information}. Another possibility is to use \emph{self-hosted}~\cite{zksyncDocs} system where users need to maintain the data for all accounts they own, which can also drastically decrease the data necessary for processing any transaction.

\point{Off-chain with on-chain settling} 
Data can be maintained by a staked committee, with misbehavior penalized through L1 slashing, ensuring data availability. 
This approach is adopted in the case of Celestia~\cite{celestia}, zkPorter~\cite{zkporter}, and Avail~\cite{avail}, as well as EigenDA~\cite{eigenda} which relies on EigenLayer's restaking~\cite{eigenlayer} to maintain data using L1 validators themselves. The overhead of replicating data in off-chain committees can be minimized by Data Availability Sampling (DAS)~\cite{DAS} where a random sampling of nodes store shards of the data. 

Among sequencer projects, Espresso introduced their DA layer called \emph{tiramisu} which combines traditional information sharding, DAS and the use of CDN networks to improve throughput~\cite{espresso2023}. Alternatively, Astria~\cite{astria} proposes leveraging Celestia's DA~\cite{celestia}.

\begin{researchquestionbox}
    How do different DA storage methods compare in terms of efficiency, cost, and security? 
\end{researchquestionbox}

\subsection{Interoperability}
\begin{center}
    \fbox{%
      \begin{minipage}{0.9\textwidth}
        \textbf{Motivation}: intra-ecosystem, globally shared \\
        \textbf{Consideration}: dertminism, data availability \\
        \textbf{Impact}: cross-rollup atomicity
      \end{minipage}
    }    
\end{center}
The rollup ecosystem is witnessing major rollups enabling developers to manage customizable rollups using their stack. This growth, however, brings to the forefront the critical issue of rollup interoperability. The primary goal is to achieve atomicity in cross-rollup transactions. In simpler terms, if two transactions span across different rollups, the ideal scenario is that both either succeed or fail. This becomes especially pertinent in the context of application-specific rollups, where there's a growing demand for aggregated liquidity in the DeFi sector.

\point{Intra-rollup} 
Currently, the pursuit of this interoperability is spearheaded by prominent rollup initiatives, notably zkSync Era Hyperscaling~\cite{zksyncDocs} and Optimism Superchain~\cite{opsuperchain}. A growing consensus in the community suggests that zk-rollups could offer a viable solution owing to their reduced latency~\cite{opsuperchain,polygon2blog}. Hyperscaling proposal outlines a mechanism where validity proofs from one rollup serve as inputs to another, facilitating seamless cross-chain transactions~\cite{polygon2blog,zksyncDocs}. This proposal also introduces an approach to aggregate these validity proofs, with the aim of reducing the associated L1 proof submission costs~\cite{zksyncDocs}. Ideally, the ambition is to bridge rollups without invoking L1, though this is primarily feasible within a singular ecosystem.

\point{Global} 
As we consider interoperability spanning multiple ecosystems, the landscape is still in its nascent stages. Present efforts, such as Astria~\cite{astria} and Espresso~\cite{espresso2023}, focus on creating a unified sequencer that can sequence transactions and relay them to any rollup. However, this unified sequencer approach must align with the sequencing protocols of the target rollup, which could introduce delays due to sequential processing. The challenge of ensuring atomicity across rollups remains an active area of exploration.
\begin{researchquestionbox}
    How can we achieve better interoperability and cross-rollup atomicity with the help of collaborative sequencing?
\end{researchquestionbox}

\subsection{Governance}
\begin{center}
    \fbox{%
      \begin{minipage}{0.9\textwidth}
        \textbf{Motivation}: no centralization, permissionless \\
        \textbf{Consideration}: Sybil resistance, incentive alignment\\
        \textbf{Impact}: collective decision
      \end{minipage}
    }    
\end{center}
In decentralized sequencer systems, updates to protocols, such as sequencing policies and reward mechanisms, are a given. These updates, with their broad implications, necessitate a consensus-driven governance approach. The power of blockchain is its collective decision-making capability. To adeptly upgrade the L1 bridge contract or adjust rollup protocols, a structured governance system is indispensable~\cite{optimismGovernance,arbitrumdocs}.
\begin{researchquestionbox}
    Which aspects of rollup sequencer operation require (decentralized) governance, in addition to upgrades?
\end{researchquestionbox}
Effective governance is pivotal for upholding fairness and sustaining decentralization. In designing these mechanisms, the emphasis on safety and security cannot be understated. Presently, projects like Optimism Mainnet, Arbitrum One, zkSync Era, and Starknet implement upgrades via multi-sig~\cite{l2beat}. This method, though superior to single-operator decisions, doesn't fully represent the diverse rollup community. A decentralized autonomous organization (DAO) structure for rollups could offer a more inclusive platform, granting every participant a say in system policies~\cite{sharma2023unpacking}. However, the inherent deliberative nature of DAOs can slow decision-making, a concern when immediate actions, like addressing security vulnerabilities, are needed. Token-based voting, given the varied expertise of participants, also poses challenges. A potential solution is a dual governance model: while global consensus issues involve all stakeholders, urgent matters could be overseen by a specialized committee of experts, elected by the DAO~\cite{coumpoundGov}. This committee, equipped with a multi-sig system, would ensure decisions are both inclusive and timely.
% \begin{researchquestionbox}
%     How can insights from established L1 governance models be adapted and integrated into L2 rollup governance structures? 
% \end{researchquestionbox}
\section{Discussions}
\label{sec:discussion}

\subsection{MEV}
\textbf{Description.} Maximal Extractable Value (MEV) is defined as the profit derived from the inclusion, exclusion, or reordering of transactions within blocks~\cite{daian2019flash}. Originally conceptualized and studied in the context of L1 blockchains~\cite{mackinga2022twap,werner2022sok}, MEV's implications extend to rollups. Sequencers, responsible for transaction ordering in rollups, have the potential to manipulate this order for their profits, such as front-running arbitrage transactions.
% The sequencer's role in generating MEV within the L2 ecosystem can be lucrative. Some proposals, like MEV auction~\cite{meva}, advocate for a bidding process for this role, granting exclusive rights for a duration to the highest bidder. Proponents believe this redirects MEV profits back into the system, while critics argue it increases user costs~\cite{mevaHarmful}. It's crucial that such mechanisms prioritize system \emph{decentralization} and ensure \emph{economic sustainability} to be considered beneficial.

% \point{Auctioning} As we shall explore more in depth later on Section~\ref{sec:discussion}, the sequencer has a predominant role for generating MEV in the layer-2 eco-system, which can be a very profitable role. Hence, proposals such as MEV auction~\cite{meva} suggest to open a bidding process for the position, guaranteeing the right to take the whole during a certain window for the party that pays the most. Those who are for this paradigm argue that it allows to redirect MEV gains back into the system, while opponents claim that it generates extra-costs for the users~\cite{mevaHarmful}. In the scope of our discussion we shall therefore highlight that such mechanisms must be carefully designed in order to not jeorpadize the \emph{decentralization} of the system and provide better \emph{economic sustainability} for the protocol in comparison with previous proposals in order to be advantageous.

The discourse surrounding MEV has elicited a range of perspectives. Some view it as a natural byproduct of a free market, while others believe it warrants attention and mitigation. Unchecked MEV in rollups poses significant challenges. Economically, it allows malicious actors to reap undue profits, potentially skewing market dynamics, especially in decentralized finance platforms. Furthermore, regular users may experience transaction delays or failures as their activities are continually overshadowed by those exploiting MEV advantages. Broadly, there are two prevailing approaches to address MEV. The first emphasizes ensuring unbiased and fair transaction ordering~\cite{cachin2022quick}. The second approach seeks to maximize the value extracted from transactions, with the intent of redistributing it among various stakeholders~\cite{mevboost}.

\point{Blind order-fairness} 
The former could be achieved by fair sequencing policy. Blind Order-Fairness serves as a cornerstone in countering Maximal Extractable Value (MEV), ensuring transactions are committed to an order without revealing their contents~\cite{malkhi2022maximal}. As detailed by Heimbach and Wattenhofer, this mechanism employs a $k$-of-$n$ secret sharing approach, where transactions are encrypted using a secret key and dispatched to validators, ensuring decryption only post blind commitment~\cite{heimbach2022sok}. Furthermore, Fino, a DAG-based BFT protocol tailored for the partial synchrony model, seamlessly integrates Blind Order-Fairness, with potential enhancements like Time-Based Fairness to further refine deterministic order and combat MEV~\cite{malkhi2022maximal,chainlinkfairseqeuencing}.

\point{Proposer-builder seperation} 
The alternative strategy to combat MEV is to distribute the profits among all stakeholders~\cite{daian2019flash}. MEV-boost~\cite{mevboost} introduces a marketplace for transaction ordering, underpinned by the Proposer Builder Separation (PBS) paradigm. PBS is a mechanism that separates the roles of proposing a block and building a block~\cite{heimbach2023ethereum}. In this design, builders identify high-profit transactions to form blocks and then bid on the potential profits for proposers. This system allows proposers to share increased profits with builders. SUAVE~\cite{suave} aims to evolve this into a marketplace where users can express their preferences and place bids. There are ongoing discussions about conducting MEV auctions using PBS on rollups~\cite{mevaHarmful,meva}. Given the distinct roles of the sequencer and executor, this system can be seen as analogous; sequencers construct the block, while executors assess its value, especially when integrated with a private mempool and encryption.

\subsection{Shared Sequencing}
% The goal of Espresso is not only to have a decentralized sequencer but also to be integrated into the blockchain stack as a sequencer shared by several rollup projects. This is not unique to Espresso, other teams have also manifested their desire to provide this service~\cite{astria}. In this section, we highlight the shortcomings of any shared sequencer project.
Projects like Astria~\cite{astria}, Espresso~\cite{espresso2023}, and Radius~\cite{radius} have been vocal in the community about the possibility of different rollups sharing the same sequencer network. In this section we shall present arguments for and against individual rollups using such a service. We see the following advantages:

\begin{itemize}
\item \textbf{Simplification.} Utilizing a shared sequencer network offers the primary benefit of negating the need for individual rollups to maintain their own infrastructure and sequencing codebase. While this advantage can be realized through a based-rollup~\cite{basedRollups}, an efficiently designed shared sequencing system holds promise for enhanced performance. 

\item \textbf{Cross-rollup inclusion.} Utilizing a shared sequencing mechanism offers the potential for users to ensure that a specific set of transactions is concurrently integrated across multiple rollup chains. This capability, when further developed, could pave the way for synchronized cross-rollup executions. 

\item \textbf{Fee unification.} A robust shared sequencer can consolidate transaction fees across platforms using one token. However, adding an intermediary between Ethereum and L2 might risk fragmenting payments and introducing an extra token.
\end{itemize}
%
% \lfnote{Con: latency}
As well as potential downsides:
\begin{itemize}
\item \textbf{Sovereignty loss.} Utilizing a shared sequencer renders client rollups reliant on an external entity beyond their control. This compromises autonomy in protocol evolution, policy decisions, and introduces challenges in managing updates or addressing bugs. Additionally, a rollup's performance and availability could be impacted by other L2s, potentially leading to sequencer congestion.

% \item \textbf{Native token concerns.} On the topic of tokens, the control of the sequencer by a rollup allows the option to open the staking of their native token for the right to take the role of the sequencer by external parties that get rewarded by MEV, which is an important mechanism for token value appreciation.

\item \textbf{MEV issues.} In the evolving domain of decentralized sequencers, ascertaining the individual contributions of each participant for MEV and devising appropriate revenue sharing mechanisms are still open challenges. Furthermore, ensuring that a shared sequencer remains neutral to the diverse MEV policies adopted by individual rollups presents a challenge. This is particularly evident when one rollup aims to maximize MEV, while another seeks its minimization or fairness.

\item \textbf{Increased latency.} In shared sequencing frameworks, both ordering and DA play key roles. However, when these tasks are executed in conjunction with the inherent sequencing protocols of individual rollups, it can introduce compounded delays. For instance, Astria's decision to utilize Celestia's DA~\cite{celestia} introduces an added latency of 12 seconds before rollup operations can be initiated~\cite{astriablog}.

% {\bf Con: Lack of atomic cross-rollup transactions.} Contrary to the shared sequencer project claims, it is impossible for the shared sequencer to guarantee that two transactions in different rollups are executed atomically. Although it is possible to guarantee that transactions are added to two rollups atomically, one of these transactions might be later rejected, with only one of them being executed.
\end{itemize}

\begin{researchquestionbox}
    How can shared sequencer respect rollups' varying sequencing policies?
\end{researchquestionbox}

\subsection{Sequencers as Provers}
In this section, we delve into the potential expansion of sequencers' responsibilities. While sequencers could serve as executors, necessitating a consensus mechanism that encompasses execution beyond mere ordering, or even as submitters, with one submitting blocks to L1, our primary focus is on the concept of sequencers taking on the role of provers. This would entail sequencers assuming a comprehensive role, potentially overseeing all system responsibilities.

\point{Starknet Proposal}
In the decentralized rollup protocol proposed by Starknet, sequencers assume the role of provers~\cite{starknetproposal}. These sequencers, by staking native tokens, actively participate in block production. A designated block proposer, chosen from the sequencers, proposes a new block. Notably, this block embeds the zk-proof of validity of the preceding sequenced block within its current header. This design effectively addresses the `runaway problem', ensuring that both execution and proof generation are central to sequencing tasks, thus preventing potential indefinite delays~\cite{starknetproposal}. Moreover, this model mitigates the buffer issue, i.e, the challenge of producing blocks that are not economically viable to prove~\cite{starknetproposal}. This alignment ensures that sequencers and provers operate with the broader system's best interests in mind.

\point{Sybil resistance}
Similarly, the Proof of Necessary Work (PoNW) introduced by Kattis and Bonneau offers a blockchain's Sybil resistance mechanism~\cite{kattis2020proof}. PoNW harnesses the energy expended in proof-of-work computations to generate succinct zk-proofs of validity. This approach ensures that the computational efforts dedicated to Sybil resistance also contribute to the system's verification. By consistently producing block headers of a uniform size, regardless of transaction volume, PoNW facilitates swift verification by stateless clients, emphasizing the benefit of rapid transaction validations~\cite{kattis2020proof}.

\point{PBS}
The B52 proposal offers a unique perspective, diverging from traditional sequencer-based systems~\cite{b52} and aligning with the Proposer Builder Separation (PBS) design paradigm. In this model, anyone can step into the role of block builders. Provers, alternatively known as proposers, are vested with the authority to select blocks and are paid fees by the sequencers (builders). In other words, provers partake in private auctions, for selling inclusion rights for a particular block. Upon observing block headers on the rollup network, these provers cast their votes in favor of the block that promises the highest profits and has the strongest likelihood of becoming the predominant chain link~\cite{b52}. Subsequently, provers generate the necessary proofs to finalize the block and ensure the builders are duly compensated~\cite{b52}. This structure mirrors the dynamics of the MEV marketplace, where a myriad of participants competitively bid for the privilege of transaction inclusion.

\point{Collusion attacks}
\label{topic:collusion-attacks}
The amalgamation of roles within a decentralized system can inadvertently engender centralization tendencies. When the duties of a sequencer overlap with those of a prover, either through collusion or by a single entity assuming both capacities, it introduces an avenue for asymmetrical advantages. Specifically, a sequencer might exhibit preferential treatment towards a particular prover by prematurely divulging details of an impending batch or by habitually selecting them for transaction validations. Such practices not only compromise the system's foundational integrity but also contravene the cardinal principles of decentralization, posing risks like reduced security and potential manipulation.

In conclusion, while integrating roles can enhance efficiency, it's crucial to balance these benefits against centralization risks and the goal of a decentralized, secure system.

\begin{researchquestionbox}
    Should there be an overlap between sequencer pools and prover pools?
\end{researchquestionbox}

\subsection{Transaction Privacy}
Transaction privacy is not only an end onto itself as a core value of many blockchain applications, but also the means to avoid censorship and to reduce MEV. We shall explore the three families of approaches we have identified in the litterature.

\point{Blinding randomness} Nazirkhanova et al.~\cite{nazirkhanova2022information} shows that a simple modification of their verifiable information dispersal protocol by adding a random row and a random column to the blocks before encoding is enough to guarantee privacy against non-colluding honest-but-curious adversaries. Though innefective against Byzantine adversaries, this approach requires no further steps after data is retrieved, having the least overhead of all solutions we shall consider.

\point{Encrypted mempools} Bebel and Ojha~\cite{ferveoMempool}, as well as Rondelet and Kilbourn~\cite{rondelet2023threshold}, among other works have explored the possibility of using encrypted mempools in order to obtain privacy. Both recognize the use of TEEs (Trusted Execution Environments) and timelock encryption as possible techniques to achieve this result, but due to the necessity of trusting the hardware manufacturer of the former and the high computational requirements of the latter, both suggest the use of threshold encryption schemes. The disadvantage of threshold encryption, however, lies in the fact that it needs extra communication at later points of the protocol in order to retrieve plain data.

\point{Secret sharing} Malkhi and Szalachowski~\cite{malkhi2022maximal} have shown that secret-sharing can be used instead of threshold encryption to achieve the same privacy results. In their work, not only do they show that secret sharing has a much lower overhead on the protocol, but also how to seamleslly integrate it into DAG-based consensus protocols.

% {\bf Unclear advantage with respect to based rollups.} It still remains unclear the advantage for a client rollup to choose a shared sequencer instead of a based rollup if the layer two does not want to maintain a sequencer network.
\section{Conclusion}
\label{sec:conclusion}
In the evolving landscape of blockchain technology, sequencers in rollups have become pivotal in enhancing scalability and efficiency. While the current generation of proposals is insightful, they only partially address the desirable properties of decentralized sequencers, often overlooking the broader system intricacies and introducing potential vulnerabilities. There is a pressing need for fair benchmarking to meet practical needs. 
We conclude that none of the designs currently being deployed are fully satisfying the requirements we have identified. Futher, they tend to focus on different, disjoint aspects of rollup design. 
Our research offers a comprehensive analysis of sequencer properties and components, paving the way for more robust sequencer designs.
\bibliographystyle{splncs04}
\bibliography{references}

\begin{thebibliography}{100}
\providecommand{\url}[1]{\texttt{#1}}
\providecommand{\urlprefix}{URL }
\providecommand{\doi}[1]{https://doi.org/#1}

\bibitem{optimismGovernance}
The optimism collective: The future of optimism governance.
  \url{https://optimism.mirror.xyz/PLrAQgE1EGRo7GRrFoztplFChnUZda4DFGW3dkQayxY},
  accessed: 2023-09-12

\bibitem{proofoffee}
0L: Proof-of-fee, part 2.
  \url{https://0l.network/2022/10/20/proof-of-fee-part-2-a-proposal/},
  accessed: 2023-09-08

\bibitem{al2019lazyledger}
Al-Bassam, M.: Lazyledger: A distributed data availability ledger with
  client-side smart contracts. arXiv preprint arXiv:1905.09274  (2019)

\bibitem{DAS}
Al{-}Bassam, M., Sonnino, A., Buterin, V.: Fraud proofs: Maximising light
  client security and scaling blockchains with dishonest majorities. CoRR
  \textbf{abs/1809.09044} (2018), \url{http://arxiv.org/abs/1809.09044}

\bibitem{amoussouguenou2019dissecting}
Amoussou-Guenou, Y., del Pozzo, A., Potop-Butucaru, M., Tucci-Piergiovanni, S.:
  Dissecting tendermint (2019)

\bibitem{arbitrumdocs}
Arbitrum: The state of arbitrum's progressive decentralization.
  \url{https://docs.arbitrum.foundation/state-of-progressive-decentralization\#3-sequencer-ownership}
  (2023), accessed: 2023-09-06

\bibitem{avail}
Avail: Avail project. \url{https://availproject.github.io/about/introduction/}
  (2023), accessed: 2023-09-10

\bibitem{b52}
Aztec: B52. \url{https://hackmd.io/VIeqkDnMScG1B-DIVIyPLg} (2023), accessed:
  2023-09-07

\bibitem{fernet}
Aztec: Sequencer selection: Fernet.
  \url{https://hackmd.io/0FwyoEjKSUiHQsmowXnJPw} (2023), accessed: 2023-09-07

\bibitem{aztecRFPResults}
{Aztec Labs}: Aztec sequencer selection finalists.
  \url{https://medium.com/aztec-protocol/aztec-sequencer-selection-finalists-122dbc6bb4}
  (2023), accessed: 2023-09-07

\bibitem{ferveoMempool}
Bebel, J., Ojha, D.: Ferveo: Threshold decryption for mempool privacy in bft
  networks. Cryptology ePrint Archive, Paper 2022/898 (2022),
  \url{https://eprint.iacr.org/2022/898},
  \url{https://eprint.iacr.org/2022/898}

\bibitem{blackshear2023sui}
Blackshear, S., Chursin, A., Danezis, G., Kichidis, A., Kokoris-Kogias, L., Li,
  X., Logan, M., Menon, A., Nowacki, T., Sonnino, A., et~al.: Sui lutris: A
  blockchain combining broadcast and consensus. Tech. rep., Technical Report.
  Mysten Labs. https://sonnino. com/papers/sui-lutris. pdf (2023)

\bibitem{lutris}
Blackshear, S., Chursin, A., Danezis, G., Kichidis, A., Kokoris-Kogias, L., Li,
  X., Logan, M., Menon, A., Nowacki, T., Sonnino, A., et~al.: Sui lutris: A
  blockchain combining broadcast and consensus. Tech. rep., Technical Report.
  Mysten Labs. https://sonnino. com/papers/sui-lutris. pdf (2023)

\bibitem{boneh2020single}
Boneh, D., Eskandarian, S., Hanzlik, L., Greco, N.: Single secret leader
  election. In: Proceedings of the 2nd ACM Conference on Advances in Financial
  Technologies. pp. 12--24 (2020)

\bibitem{astria}
Bowen, J., Oroshiba, J.: Introducing astria: The shared sequencer network.
  \url{https://blog.astria.org/introducing-astria/} (04 2023), accessed:
  2023-08-02

\bibitem{buchman2016tendermint}
Buchman, E.: Tendermint: Byzantine fault tolerance in the age of blockchains.
  Ph.D. thesis, University of Guelph (2016)

\bibitem{buchman2022revisiting}
Buchman, E., Guerraoui, R., Komatovic, J., Milosevic, Z., Seredinschi, D.A.,
  Widder, J.: Revisiting tendermint: Design tradeoffs, accountability, and
  practical use. In: 2022 52nd Annual IEEE/IFIP International Conference on
  Dependable Systems and Networks-Supplemental Volume (DSN-S). pp. 11--14. IEEE
  (2022)

\bibitem{buchman2018latest}
Buchman, E., Kwon, J., Milosevic, Z.: The latest gossip on bft consensus. arXiv
  preprint arXiv:1807.04938  (2018)

\bibitem{bunz2017proofs}
B{\"u}nz, B., Goldfeder, S., Bonneau, J.: Proofs-of-delay and randomness
  beacons in ethereum. IEEE Security and Privacy on the blockchain (IEEE S\&B)
  (2017)

\bibitem{overloadEthereum}
Buterin, V.: Don't overload ethereum's consensus.
  \url{https://vitalik.ca/general/2023/05/21/dont_overload.html} (2023),
  accessed: 2023-09-12

\bibitem{cachin2022quick}
Cachin, C., Mi{\'c}i{\'c}, J., Steinhauer, N., Zanolini, L.: Quick order
  fairness. In: International Conference on Financial Cryptography and Data
  Security. pp. 316--333. Springer (2022)

\bibitem{castro1999practical}
Castro, M., Liskov, B., et~al.: Practical byzantine fault tolerance. In: OsDI.
  vol.~99, pp. 173--186 (1999)

\bibitem{catalini2020some}
Catalini, C., Gans, J.S.: Some simple economics of the blockchain.
  Communications of the ACM  \textbf{63}(7),  80--90 (2020)

\bibitem{celestia}
Celestia: Celestia's data availability layer.
  \url{https://docs.celestia.org/concepts/how-celestia-works/data-availability-layer/}
  (2023), accessed: 2023-09-10

\bibitem{chan2020streamlet}
Chan, B.Y., Shi, E.: Streamlet: Textbook streamlined blockchains. In:
  Proceedings of the 2nd ACM Conference on Advances in Financial Technologies.
  pp. 1--11 (2020)

\bibitem{proofofgovernance}
Charbonneau, J.: Endgame: Proof of governance.
  \url{https://medium.com/@wunderlichvalentin/endgame-proof-of-governance-6b792179c001}
  (2023), accessed: 2023-09-07

\bibitem{chaudhry2018consensus}
Chaudhry, N., Yousaf, M.M.: Consensus algorithms in blockchain: Comparative
  analysis, challenges and opportunities. In: 2018 12th International
  Conference on Open Source Systems and Technologies (ICOSST). pp. 54--63. IEEE
  (2018)

\bibitem{cohen2019chia}
Cohen, B., Pietrzak, K.: The chia network blockchain. White Paper, Chia. net
  \textbf{9} (2019)

\bibitem{cohen2022aware}
Cohen, S., Gelashvili, R., Kogias, L.K., Li, Z., Malkhi, D., Sonnino, A.,
  Spiegelman, A.: Be aware of your leaders. In: International Conference on
  Financial Cryptography and Data Security. pp. 279--295. Springer (2022)

\bibitem{corso2019performance}
Corso, A.: Performance analysis of proof-of-elapsed-time (poet) consensus in
  the sawtooth blockchain framework. Ph.D. thesis, University of Oregon (2019)

\bibitem{rollupeconomics}
Crapis, D.: Rollups are real — rollup economics 2.0.
  \url{https://davidecrapis.notion.site/Rollups-are-Real-Rollup-Economics-2-0-2516079f62a745b598133a101ba5a3de},
  accessed: 2023-09-08

\bibitem{daian2019flash}
Daian, P., Goldfeder, S., Kell, T., Li, Y., Zhao, X., Bentov, I., Breidenbach,
  L., Juels, A.: Flash boys 2.0: Frontrunning, transaction reordering, and
  consensus instability in decentralized exchanges. arXiv preprint
  arXiv:1904.05234  (2019)

\bibitem{narwhal}
Danezis, G., Kokoris-Kogias, L., Sonnino, A., Spiegelman, A.: Narwhal and tusk:
  a dag-based mempool and efficient bft consensus. In: Proceedings of the
  Seventeenth European Conference on Computer Systems. pp. 34--50 (2022)

\bibitem{danksharding}
Dankrad: New sharding design with tight beacon and shard block integration.
  \url{https://notes.ethereum.org/@dankrad/new\_sharding} (2023), accessed:
  2023-09-10

\bibitem{basedRollups}
Drake, J.: Based rollups—superpowers from l1 sequencing.
  \url{https://ethresear.ch/t/based-rollups-superpowers-from-l1-sequencing/15016}
  (2023), accessed: 2023-09-07

\bibitem{eigenlayer}
{Eigen Layer}: Eigenlayer: The restaking collective.
  \url{https://www.eigenlayer.xyz/}, accessed: 2023-09-06

\bibitem{eigenda}
{EigenLabs}: Intro to {EigenDA}: Hyperscale data availability for rollups.
  \url{https://www.blog.eigenlayer.xyz/intro-to-eigenda-hyperscale-data-availability-for-rollups/}
  (2023), accessed: 2023-09-10

\bibitem{astriablog}
Eshita: Introducing the astria development cluster.
  \url{https://blog.astria.org/introducing-the-astria-development-cluster/},
  accessed: 2023-08-07

\bibitem{espresso2023}
{Espresso Systems}: The espresso sequencer: Hotshot consensus and tiramisu data
  availability.
  \url{https://github.com/EspressoSystems/HotShot/blob/main/docs/espresso-sequencer-paper.pdf},
  accessed: 2023-08-06

\bibitem{ethereumLeader}
{Ethereum Foundation}: Ethereum beacon-chain.
  \url{https://github.com/ethereum/annotated-spec/blob/master/phase0/beacon-chain.md\#aside-randao-seeds-and-committee-generation}
  (2023), accessed: 2023-09-07

\bibitem{dankshardingBlog}
{Ethereum Foundation}: Ethereum roadmap: Danksharding.
  \url{https://ethereum.org/en/roadmap/danksharding/} (2023), accessed:
  2023-09-10

\bibitem{ethereumSSLE}
{Ethereum Foundation}: Ethereum roadmap: Secret leader election.
  \url{https://ethereum.org/en/roadmap/secret-leader-election/} (2023),
  accessed: 2023-09-07

\bibitem{fanti2019economics}
Fanti, G., Kogan, L., Viswanath, P.: Economics of proof-of-stake payment
  systems. In: Working paper (2019)

\bibitem{mevaHarmful}
Felten, E.: Mev auctions considered harmful.
  \url{https://medium.com/offchainlabs/mev-auctions-considered-harmful-fa72f61a40ea}
  (2023), accessed: 2023-09-10

\bibitem{suave}
Flashbots: The future of mev is suave.
  \url{https://notes.ethereum.org/@vbuterin/pbs_censorship_resistance},
  accessed: 2023-09-15

\bibitem{meva}
Floersch, K.: Mev auction: Auctioning transaction ordering rights as a solution
  to miner extractable value.
  \url{https://ethresear.ch/t/mev-auction-auctioning-transaction-ordering-rights-as-a-solution-to-miner-extractable-value/6788}
  (2023), accessed: 2023-09-10

\bibitem{garay2015bitcoin}
Garay, J., Kiayias, A., Leonardos, N.: The bitcoin backbone protocol: Analysis
  and applications. In: Annual international conference on the theory and
  applications of cryptographic techniques. pp. 281--310. Springer (2015)

\bibitem{gavzi2023fait}
Ga{\v{z}}i, P., Kiayias, A., Russell, A.: Fait accompli committee selection:
  Improving the size-security tradeoff of stake-based committees. Cryptology
  ePrint Archive  (2023)

\bibitem{gilad2017algorand}
Gilad, Y., Hemo, R., Micali, S., Vlachos, G., Zeldovich, N.: Algorand: Scaling
  byzantine agreements for cryptocurrencies. In: Proceedings of the 26th
  symposium on operating systems principles. pp. 51--68 (2017)

\bibitem{goldreich1991proofs}
Goldreich, O., Micali, S., Wigderson, A.: Proofs that yield nothing but their
  validity or all languages in np have zero-knowledge proof systems. Journal of
  the ACM (JACM)  \textbf{38}(3),  690--728 (1991)

\bibitem{tezos}
Goodman, L.: Tezos — a self-amending crypto-ledger white paper.
  \url{https://tezos.com/whitepaper.pdf}, accessed: 2023-06-28

\bibitem{gorzny2022ideal}
Gorzny, J., Po-An, L., Derka, M.: Ideal properties of rollup escape hatches.
  In: Proceedings of the 3rd International Workshop on Distributed
  Infrastructure for the Common Good. pp. 7--12 (2022)

\bibitem{gupta2019depth}
Gupta, S., Hellings, J., Rahnama, S., Sadoghi, M.: An in-depth look of bft
  consensus in blockchain: Challenges and opportunities. In: Proceedings of the
  20th international middleware conference tutorials. pp. 6--10 (2019)

\bibitem{mevboost}
Hasu, Gosselin, S.: Why run mev-boost?
  \url{https://writings.flashbots.net/why-run-mevboost/}, accessed: 2023-09-15

\bibitem{heimbach2023ethereum}
Heimbach, L., Kiffer, L., Torres, C.F., Wattenhofer, R.: Ethereum's
  proposer-builder separation: Promises and realities. arXiv preprint
  arXiv:2305.19037  (2023)

\bibitem{heimbach2022sok}
Heimbach, L., Wattenhofer, R.: Sok: Preventing transaction reordering
  manipulations in decentralized finance. arXiv preprint arXiv:2203.11520
  (2022)

\bibitem{chainlinkfairseqeuencing}
Juels, A.: Fair sequencing services: Enabling a provably fair defi ecosystem.
  \url{https://blog.chain.link/chainlink-fair-sequencing-services-enabling-a-provably-fair-defi-ecosystem/}
  (2023), accessed: 2023-09-12

\bibitem{kalodner2018arbitrum}
Kalodner, H., Goldfeder, S., Chen, X., Weinberg, S.M., Felten, E.W.: Arbitrum:
  Scalable, private smart contracts. In: 27th USENIX Security Symposium (USENIX
  Security 18). pp. 1353--1370 (2018)

\bibitem{kattis2020proof}
Kattis, A., Bonneau, J.: Proof of necessary work: Succinct state verification
  with fairness guarantees. Cryptology ePrint Archive  (2020)

\bibitem{allYouNeedIsDAG}
Keidar, I., Kokoris-Kogias, E., Naor, O., Spiegelman, A.: All you need is dag.
  In: Proceedings of the 2021 ACM Symposium on Principles of Distributed
  Computing. pp. 165--175 (2021)

\bibitem{kelkar2021themis}
Kelkar, M., Deb, S., Long, S., Juels, A., Kannan, S.: Themis: Fast, strong
  order-fairness in byzantine consensus. Cryptology ePrint Archive  (2021)

\bibitem{kelkar2020order}
Kelkar, M., Zhang, F., Goldfeder, S., Juels, A.: Order-fairness for byzantine
  consensus. In: Advances in Cryptology--CRYPTO 2020: 40th Annual International
  Cryptology Conference, CRYPTO 2020, Santa Barbara, CA, USA, August 17--21,
  2020, Proceedings, Part III 40. pp. 451--480. Springer (2020)

\bibitem{whisky}
Kunz, C.: Whisk-y: should we use whisk for sequencer selection?
  \url{https://discourse.aztec.network/t/whisk-y-should-we-use-whisk-for-sequencer-selection/365}
  (2023), accessed: 2023-09-07

\bibitem{kursawe2020wendy}
Kursawe, K.: Wendy, the good little fairness widget: Achieving order fairness
  for blockchains. In: Proceedings of the 2nd ACM Conference on Advances in
  Financial Technologies. pp. 25--36 (2020)

\bibitem{kwon2014tendermint}
Kwon, J.: Tendermint: Consensus without mining. Draft v. 0.6, fall
  \textbf{1}(11),  1--11 (2014)

\bibitem{cosmos}
Kwon, J., Buchman, E.: Cosmos whitepaper.
  \url{https://v1.cosmos.network/resources/whitepaper}, accessed: 2023-06-28

\bibitem{coumpoundGov}
Leshner, R.: Set pause guardian to community multi-sig.
  \url{https://compound.finance/governance/proposals/57}, accessed: 2023-09-18

\bibitem{mackinga2022twap}
Mackinga, T., Nadahalli, T., Wattenhofer, R.: Twap oracle attacks: Easier done
  than said? In: 2022 IEEE International Conference on Blockchain and
  Cryptocurrency (ICBC). pp.~1--8. IEEE (2022)

\bibitem{hotstuff2}
Malkhi, D., Nayak, K.: Extended abstract: Hotstuff-2: Optimal two-phase
  responsive bft. Cryptology ePrint Archive, Paper 2023/397 (2023),
  \url{https://eprint.iacr.org/2023/397}

\bibitem{malkhi2022maximal}
Malkhi, D., Szalachowski, P.: Maximal extractable value (mev) protection on a
  dag. arXiv preprint arXiv:2208.00940  (2022)

\bibitem{malkhi2023lessons}
Malkhi, D., Yin, M.: Lessons from hotstuff. In: Proceedings of the 5th workshop
  on Advanced tools, programming languages, and PLatforms for Implementing and
  Evaluating algorithms for Distributed systems. pp.~1--8 (2023)

\bibitem{lessonsHotstuff}
Malkhi, D., Yin, M.: Lessons from hotstuff. In: Proceedings of the 5th workshop
  on Advanced tools, programming languages, and PLatforms for Implementing and
  Evaluating algorithms for Distributed systems. pp.~1--8 (2023)

\bibitem{mamageishvili2022efficient}
Mamageishvili, A., Felten, E.W.: Efficient l2 batch posting strategy on l1.
  arXiv preprint arXiv:2212.10337  (2022)

\bibitem{mamageishvili2023buying}
Mamageishvili, A., Kelkar, M., Schlegel, J.C., Felten, E.W.: Buying time:
  Latency racing vs. bidding in fair transaction ordering. arXiv preprint
  arXiv:2306.02179  (2023)

\bibitem{boldarbitrum}
Mario~M., A., Lee, B., Buckland, C., Yafah, E., Edward, W.F., Daniel, G., Raul,
  J., Mahimna, K., Harry, N., Terence, T., Preston, V.L.: Bold: Bounded
  liquidity delay in a rollup challenge protocol.
  \url{https://github.com/OffchainLabs/bold/blob/main/docs/research-specs/BOLDChallengeProtocol.pdf},
  accessed: 2023-09-10

\bibitem{zkporter}
{Matter Labs}: zkporter: a breakthrough in l2 scaling.
  \url{https://blog.matter-labs.io/zkporter-a-breakthrough-in-l2-scaling-ed5e48842fbf}
  (2023), accessed: 2023-09-10

\bibitem{mccorry2021sok}
McCorry, P., Buckland, C., Yee, B., Song, D.: Sok: Validating bridges as a
  scaling solution for blockchains. Cryptology ePrint Archive  (2021)

\bibitem{micali1999verifiable}
Micali, S., Rabin, M., Vadhan, S.: Verifiable random functions. In: 40th annual
  symposium on foundations of computer science (cat. No. 99CB37039). pp.
  120--130. IEEE (1999)

\bibitem{motepalli2021reward}
Motepalli, S., Jacobsen, H.A.: Reward mechanism for blockchains using
  evolutionary game theory. In: 2021 3rd Conference on Blockchain Research \&
  Applications for Innovative Networks and Services (BRAINS). pp. 217--224.
  IEEE (2021)

\bibitem{motepalli2022decentralizing}
Motepalli, S., Jacobsen, H.A.: Decentralizing permissioned blockchain with
  delay towers. arXiv preprint arXiv:2203.09714  (2022)

\bibitem{10237020}
Motepalli, S., Jacobsen, H.A.: Analyzing geospatial distribution in
  blockchains. In: 2023 IEEE International Conference on Decentralized
  Applications and Infrastructures (DAPPS). pp. 100--108 (2023).
  \doi{10.1109/DAPPS57946.2023.00022}

\bibitem{sui}
{MystenLabs Team}: The sui smart contracts platform (2022),
  \url{https://docs.sui.io/paper/sui.pdf}

\bibitem{nakamoto2008bitcoin}
Nakamoto, S.: Bitcoin: A peer-to-peer electronic cash system. Decentralized
  business review  (2008)

\bibitem{nazirkhanova2022information}
Nazirkhanova, K., Neu, J., Tse, D.: Information dispersal with provable
  retrievability for rollups. In: Proceedings of the 4th ACM Conference on
  Advances in Financial Technologies. pp. 180--197 (2022)

\bibitem{opsuperchain}
Optimism: Op stack docs - superchain explainer.
  \url{https://stack.optimism.io/docs/understand/explainer/}, accessed:
  2023-08-08

\bibitem{polygondata}
Polygon: Staking all validators.
  \url{https://staking.polygon.technology/validators}, accessed: 2023-06-28

\bibitem{polygon2blog}
Polygon: Polygon 2.0: Protocol architecture.
  \url{https://polygon.technology/blog/polygon-2-0-protocol-vision-and-architecture}
  (2023), accessed: 2023-09-08

\bibitem{radius}
Radius: Introduction to radius.
  \url{https://docs.theradius.xyz/overview/introduction-to-radius}, accessed:
  2023-09-12

\bibitem{rondelet2023threshold}
Rondelet, A., Kilbourn, Q.: Threshold encrypted mempools: Limitations and
  considerations (2023)

\bibitem{saleh2021blockchain}
Saleh, F.: Blockchain without waste: Proof-of-stake. The Review of financial
  studies  \textbf{34}(3),  1156--1190 (2021)

\bibitem{sharma2023unpacking}
Sharma, T., Kwon, Y., Pongmala, K., Wang, H., Miller, A., Song, D., Wang, Y.:
  Unpacking how decentralized autonomous organizations (daos) work in practice.
  arXiv preprint arXiv:2304.09822  (2023)

\bibitem{shoal}
Spiegelman, A., Aurn, B., Gelashvili, R., Li, Z.: Shoal: Improving dag-bft
  latency and robustness. arXiv preprint arXiv:2306.03058  (2023)

\bibitem{bullshark}
Spiegelman, A., Giridharan, N., Sonnino, A., Kokoris-Kogias, L.: Bullshark: Dag
  bft protocols made practical. In: Proceedings of the 2022 ACM SIGSAC
  Conference on Computer and Communications Security. pp. 2705--2718 (2022)

\bibitem{bbca}
Stathakopoulou, C., Wei, M., Yin, M., Zhang, H., Malkhi, D.: Bbca-ledger: High
  throughput consensus meets low latency. arXiv preprint arXiv:2306.14757
  (2023)

\bibitem{l2beat}
research team, L.: L2 beat: The state of layer two ecosystem.
  \url{https://l2beat.com/scaling/summary} (2023), accessed: 2023-08-26

\bibitem{tsabary2018gap}
Tsabary, I., Eyal, I.: The gap game. In: Proceedings of the 2018 ACM SIGSAC
  conference on Computer and Communications Security. pp. 713--728 (2018)

\bibitem{starknetTendermint}
Volokh, I.: Tendermint for starknet.
  \url{https://community.starknet.io/t/tendermint-for-starknet/98248},
  accessed: 2023-06-28

\bibitem{starknetproposal}
Volokh, I.: Starknet simple decentralized protocol proposal.
  \url{https://community.starknet.io/t/simple-decentralized-protocol-proposal/99693}
  (2023), accessed: 2023-09-06

\bibitem{werner2022sok}
Werner, S., Perez, D., Gudgeon, L., Klages-Mundt, A., Harz, D., Knottenbelt,
  W.: Sok: Decentralized finance (defi). In: Proceedings of the 4th ACM
  Conference on Advances in Financial Technologies. pp. 30--46 (2022)

\bibitem{yin2019hotstuff}
Yin, M., Malkhi, D., Reiter, M.K., Gueta, G.G., Abraham, I.: Hotstuff: Bft
  consensus with linearity and responsiveness. In: Proceedings of the 2019 ACM
  Symposium on Principles of Distributed Computing. pp. 347--356 (2019)

\bibitem{zhang2022reaching}
Zhang, G., Pan, F., Dang'ana, M., Mao, Y., Motepalli, S., Zhang, S., Jacobsen,
  H.A.: Reaching consensus in the byzantine empire: A comprehensive review of
  bft consensus algorithms. arXiv preprint arXiv:2204.03181  (2022)

\bibitem{zhang2020byzantine}
Zhang, Y., Setty, S., Chen, Q., Zhou, L., Alvisi, L.: Byzantine ordered
  consensus without byzantine oligarchy. In: 14th USENIX Symposium on Operating
  Systems Design and Implementation (OSDI 20). pp. 633--649 (2020)

\bibitem{zksyncDocs}
zkSync: Technical reference. \url{https://era.zksync.io/docs/reference/},
  accessed: 2023-08-10

\end{thebibliography}

\appendix
\label{appendix:starknet}
\section{Applying Our Framework on Starknet's Decentralized Protocol Proposal}
In this section, we employ our framework to critically assess the Starknet decentralized protocol proposal~\cite{starknetproposal,starknetTendermint}, which outlines a vision for decentralizing zk-rollups illustrated in Section~\ref{sec:workflow}. Articulated through a sequence of blog posts on their forum, the Starknet proposal is distinguished by its completeness, even though it overlooks certain facets. Our objective is to gain an in-depth insight into Starknet's approach, bearing in mind its nascent stage and absence of full-fledged implementations.

At the heart of the Starknet proposal lie the principles of Proof of Stake (PoS), the Tendermint consensus algorithm~\cite{buchman2016tendermint,kwon2014tendermint}, and the assignment of prover tasks to sequencers, see Section~\ref{topic:collusion-attacks}. Equipped with this foundational knowledge, we will commence by exploring the distinct components, followed by an evaluation of their congruence with the ideal properties of sequencers.

\subsection{Components}

\point{Committee Selection}
Starknet incorporates Proof of Stake (PoS) using its native tokens as a mechanism for Sybil resistance. The core of the committee selection process is anchored in an L1 smart contract. This contract is multifunctional, overseeing deposits, withdrawals, and notably, the selection of proposers. By anchoring its operations on an L1 smart contract, Starknet appears to harness the security guarantees of L1, ensuring a transparent and robust committee selection process.

\point{Sequencing Policy}
The Starknet proposal does not explicitly detail its sequencing policy. Given the assumption of rational sequencers, it is plausible that maximizing value extraction from pending transactions becomes the de-facto approach. Intriguingly, with sequencers doubling as provers, Starknet seems to inherently address the buffer issue, i.e, challenge of sequencing blocks which are not economically viable to prove. This implies that sequencers would be disinclined to produce blocks that are not economically viable to prove.

\point{Consensus Mechanism}
Initially, Starknet's proposal leaned towards DAG-based designs~\cite{bullshark}. However, a shift towards Tendermint is evident in recent discussions, primarily due to its simplicity and proven reliability. Starknet emphasizes state machine replication over mere sequencing and incorporates periodic checkpointing to ensure safety amidst asynchrony, aiming for swift finality. 

Starknet made adaptations to Tendermint. A notable modification is addressing the ``runaway" scenario, where provers become a bottleneck. To address this, the proposal suggests embedding the proof of a previously proposed block within the header of the succeeding block. It is worth noting that Tendermint's performance might lag behind DAG-based designs in sequencing contexts. However, this is not a challenge for Starknet because the generating proofs are the critical component in this system. 

\point{Proposer Selection}
Proposers are chosen through L1 smart contact based on a combination of inherent randomness and the stake they've deposited. The leader election function operates by mapping each user's stake to a specific interval, subsequently merging these intervals, and then selecting a point at random. This method guarantees the random and exclusive selection of a single leader.

\point{Reward Mechanism}
The Starknet proposal remains silent on the specifics of the reward mechanism. However, it does introduce slashing as a countermeasure to the ``nothing at stake" problem. By anchoring the slashing logic on L1, Starknet aims to fortify its defenses against equivocation, even when confronted with a majority of malicious stakeholders within the rollup.

\point{Data Availability}
The Starknet proposal does not explicitly address Data Availability (DA). However, DA can be inferred from the use of the Tendermint consensus. All sequencers within the system retain transactional data on the rollup. Additionally, the practice of periodic checkpointing serves to reinforce and ensure data availability.

\point{Interoperability}
The Starknet proposal does not delve into the specifics of interoperability, especially concerning interactions with other rollups, be it within the Starknet ecosystem or external to it. This aspect remains an area of exploration and potential development.

\point{Governance}
The governance structure within Starknet's framework is not explicitly detailed. However, there is a reference to the possibility of implementing social recovery mechanisms via L1 smart contracts, suggesting a direction towards decentralized governance and recovery solutions.

\subsection{Ideal Properties}
\subsubsection{Basic functionality}
\begin{itemize}
    \item \textbf{Ordering transactions:} Starknet's proposal addresses the sequencing of transactions and the process of block generation.
    \item \textbf{Validity:} The integration of state machine replication alongside the consensus mechanism ensures the correctness of each transaction.
    \item \textbf{Liveness:} Starknet assures system liveness under the premise of partial synchrony. While there are propositions to ensure liveness in asynchronous settings, these might compromise the system's safety guarantees.
    \item \textbf{Resilience:} The proposal adopts a quorum-based consensus approach, which is resilient against adversarial behavior. Specifically, the system can withstand up to \(f\) faulty sequencers in a network of \(3f+1\) sequencers.
\end{itemize}

\subsubsection{Decentralization}
\begin{itemize}
\item \textbf{No Centralization:} Starknet operates without a centralized authority. Instead, a committee, selected on L1, oversees the rollup's management. However, the degree of decentralization is directly proportional to the distribution of stake.
\item \textbf{Permissionless:} The system is open to any staker wishing to participate as a sequencer or prover, facilitated through the L1 contract. This openness is underpinned by the security guarantees of L1.
\item \textbf{Collective Decision-Making:} Sequencers achieve consensus through a super-majority quorum, ensuring collective decision-making.
\item \textbf{Sybil Resistance:} The protocol leverages PoS to counteract Sybil attacks, ensuring that participants cannot flood the network with malicious nodes.
\item \textbf{Geospatial Distribution:} The proposal does not delve into geospatial distribution, indicating potential vulnerabilities related to geospatial centralization. The current design might also be susceptible to latency racing issues.
\end{itemize}

\subsubsection{Performance}

\begin{itemize}
    \item \textbf{High Throughput:} Leveraging Tendermint, Starknet achieves a throughput surpassing Ethereum. However, it might still lag behind latest designs or DAG-based systems.
    \item \textbf{Low Latency:} Owing to the Tendermint consensus, Starknet's latency is superior to Ethereum's. Nonetheless, the practice of embedding the proof of a previously proposed block in the current header could introduce sequencing delays.
    \item \textbf{Fast Finality:} Finality is attained once checkpoints are committed to L1. This process, encompassing proof generation and L1 submission, could span several minutes, if not longer.
    \item \textbf{Minimal Computational Resources:} The protocol's reliance on a PoS quorum-based consensus ensures that computational demands for sequencing remain minimal.
\end{itemize}

\subsubsection{Safety}

\begin{itemize}
    \item \textbf{Byzantine Fault Tolerance:} The system is designed to withstand up to \(f\) Byzantine faults in a network of \(3f+1\) nodes. Safety gurantees are provided under conditions of partial synchrony.
    
    \item \textbf{Data Availability:} Consensus regarding data availability and transaction execution is achieved for every transaction. This, however, necessitates trust in the rollup committee. Additionally, periodic checkpointing ensures DA commitment on L1.
    
    \item \textbf{DDoS Resistance:} While not explicitly mentioned in the proposal, the protocol could potentially leverage the resilience of full nodes and other strategies that have proven effective in L1 environments.
\end{itemize}

\subsubsection{Fairness}

\begin{itemize}
    \item \textbf{Deterministic Sequencing:} Sequencing in the proposal is non-deterministic. Given the current specifications, proposers possess the capability to front-run or delay transactions.
    
    \item \textbf{Censorship Resistance:} The rotation of block proposers ensures that every transaction is eventually included, bolstering censorship resistance.
    
    \item \textbf{Transparency:} Processes for committee selection, proposer selection, and sequencing are transparent. However, the power vested in the proposer to selectively include or exclude transactions can potentially compromise this transparency.
    
    \item \textbf{Non-discriminatory Fees:} The proposal does not delve into the specifics of the fee structure.
\end{itemize}

\subsubsection{Economic Sustainability}

\begin{itemize}
    \item \textbf{Cost Efficiency:} The proposal does not provide details on cost efficiency.
    
    \item \textbf{Incentive Alignment:} The proposal alludes to rewarding provers. Given the overlap between sequencers and provers, this is a promising direction. However, this could lead to collusion attacks, see Section~\ref{topic:collusion-attacks}. The specific mechanisms are not elaborated upon.
    
    \item \textbf{Reward Distribution:} The proposal does not discuss reward distribution, except for the aforementioned potential rewards for provers.
    
    \item \textbf{Slashing Mechanisms:} The proposal introduces the possibility of slashing sequencers for malicious actions and addresses the ``nothing at stake" problem.
\end{itemize}

\subsubsection{Interoperability}

\begin{itemize}
    \item \textbf{Intra-ecosystem Sequencing:} The proposal does not discuss intra-ecosystem sequencing.
    
    \item \textbf{Global Shared Sequencing:} The proposal does not address global shared sequencing.
    
    \item \textbf{Atomic Composability:} The proposal does not touch upon atomic composability.
\end{itemize}

\end{document}